\documentclass[referee,sn-standardnature]{sn-jnl}

\usepackage{amsmath}
\usepackage{subcaption}
\usepackage{dblfloatfix} 
\usepackage{afterpage}
\usepackage{float}
\usepackage[normalem]{ulem}  

\newcommand{\gatex}{\texttt{X}}
\newcommand{\gatey}{\texttt{Y}}
\newcommand{\gatez}{\texttt{Z}}
\newcommand{\gateh}{\texttt{H}}

\newcommand{\gatet}{\texttt{T}}
\newcommand{\gatecx}{\texttt{CX}}
\newcommand{\gatecz}{\texttt{CZ}}
\newcommand{\cirq}{\texttt{Cirq}}

\jyear{2023}%

\begin{document}

\title[Quantum circuit fidelity estimation using machine learning]{Quantum circuit fidelity estimation using machine learning}


\author*[1,2]{\fnm{Avi} \sur{Vadali}}\email{avadali@caltech.edu}

\author*[3,4]{\fnm{Rutuja} \sur{Kshirsagar}}\email{rkshirsagar@fujitsu.com}

\author[5]{\fnm{Prasanth} \sur{Shyamsundar}}

\author[5]{\fnm{Gabriel N.} \sur{Perdue}}

\affil[1]{\orgname{The Latin School of Chicago}, \orgaddress{\street{59 W North Blvd}, \city{Chicago}, \postcode{60610}, \state{IL}, \country{USA}}}

\affil*[2]{\orgname{California Institute of Technology}, \orgaddress{\street{1200 E California Blvd}, \city{Pasadena}, \postcode{91125}, \state{CA}, \country{USA}}}

\affil[3]{\orgname{Virginia Polytechnic Institute and State University}, \orgaddress{\city{Blacksburg}, \postcode{24061-0002}, \state{VA}, \country{USA}}}

\affil*[4]{\orgname{Fujitsu Research of America, Inc.}, \orgaddress{\street{350 Cobalt Way}, \city{Sunnyvale}, \postcode{94085}, \state{CA}, \country{USA}}}

\affil[5]{\orgdiv{Fermilab Quantum Institute}, \orgname{Fermi National Accelerator Laboratory}, \orgaddress{\street{PO Box 500}, \city{Batavia}, \postcode{60510-0500}, \state{IL}, \country{USA}}}

\abstract{
  The computational power of real-world quantum computers is limited by errors. When using quantum computers to perform algorithms which cannot be efficiently simulated classically, it is important to quantify the accuracy with which the computation has been performed. In this work we introduce a machine-learning-based technique to estimate the fidelity between the state produced by a noisy quantum circuit and the target state corresponding to ideal noise-free computation.
  Our machine learning model is trained in a supervised manner, using smaller or simpler circuits for which the fidelity can be estimated using other techniques like direct fidelity estimation and quantum state tomography.
  We demonstrate that, for simulated random quantum circuits with a realistic noise model, the trained model can predict the fidelities of more complicated circuits for which such methods are infeasible.
  In particular, we show the trained model may make predictions for circuits with higher degrees of entanglement than were available in the training set, and that the model may make predictions for non-Clifford circuits even when the training set included only Clifford-reducible circuits.
  This empirical demonstration suggests classical machine learning may be useful for making predictions about beyond-classical quantum circuits for some non-trivial problems.

}



\maketitle

\section{Introduction} 
\label{sec:introduction}

Estimating the quality of computations performed by a noisy quantum computer is an important task.
It allows us to benchmark current and future quantum computers and calculate the credibility of their computations \cite{Proctor2022EstablishingTI}. 
Furthermore, knowledge of the characteristics of various different circuit implementations of a given quantum algorithm can inform optimal implementation in a noise-aware manner.
Computation quality may be quantified using the fidelity between the state produced by the physical circuit and the target state corresponding to an ideal (noiseless) circuit.
Henceforth, we will refer to this fidelity (assuming that all the qubits are initialized to $\vert0\rangle$ prior to the computation) simply as the fidelity of the circuit.
Several techniques exist in the literature to estimate this fidelity, e.g., quantum state tomography \cite{PhysRevLett.78.390} and direct fidelity estimation (DFE) \cite{PhysRevLett.106.230501,PhysRevLett.107.210404}. 
Mirror Circuit Fidelity Estimation (MCFE) is a technique for estimating the entanglement or process fidelity, which is related, but not identical, to state fidelity considered in this paper \cite{Proctor2022EstablishingTI}.

The available techniques vary both in a) situations where they are applicable, and b) their classical and quantum computational cost.
Techniques like quantum state tomography and direct fidelity estimation can accurately estimate fidelities for all noise models.
However, they require access to the target quantum state from classical simulations and they involve running the circuit under consideration multiple times on the quantum hardware.
The classical and quantum computational costs required for these techniques can be prohibitively expensive for large circuits.

The MCFE technique does not require access to the target state from simulations.
It is compatible with a wide range of noise models and may even account for calibration errors in the implementation of the gates that are exactly cancelled by their inverses through the insertion of a random Pauli layer between the forward and inverse circuits \cite{Proctor2022}.
However, it still requires one to run multiple randomly chosen variations of the relevant circuit on the quantum hardware at inference time. 

In this paper, we are interested in fidelity estimation techniques which do not require running the relevant circuits on a quantum computer.
Such techniques rely upon learning the noise-characteristics of the quantum device in advance through appropriate experiments, and subsequently using this to predict the fidelities of new circuits.
For notational convenience, we will refer to such techniques as ``fully-classical'' fidelity estimation techniques, despite the initial learning phase, which requires quantum hardware.
An example of such a technique is estimating the fidelity of a circuit as the product of fidelities of all the gates used in the circuit, which are measured ahead of time during calibration stages --- a common ``back of the envelope'' method employed by practitioners.
Fully-classical fidelity estimation is useful for, among other things, to perform an initial evaluation of the potential circuit implementations of a quantum algorithm, for example during circuit compilation.
This allows a user to narrow the focus down to a smaller number of candidate circuit implementations that may be further refined with more sophisticated, albeit more computationally expensive, techniques like cross-entropy benchmarking (XEB) calibration \cite{Boixo2018}.

However, fully-classical fidelity estimation is challenging to perform accurately. 
The difficulty --- and poor performance --- of this sort of fidelity estimation arises from the difficult-to-model nature of noise in quantum devices.
For example, it is well known that the product of gate fidelities is a weak predictor of true circuit fidelities.
Machie learning (ML) algorithms offer prosmise as a heuristic family of models capable of implicitly fitting a parameterized model of the performance of a quantum computer and potentially offering stronger fidelity predictors.

One of the main challenges in deploying ML for the purpose of fidelity estimation is developing a training dataset because accurately estimating the true fidelity of high qubit-count circuits is prohibitively expensive.
An interesting question is whether it is possible to train an effective ML algorithm for fidelity prediction using small, easy to simulate circuits, and yet run inference on large, difficult to simulate circuits.
We know fully simulating large quantum circuits exactly is a classically hard problem with exponential scaling, but it is an open problem as to whether heuristic models may learn to predict projections of the information encoded in full quantum circuit evolution to arbitrary accuracy.

In this work we propose an approach whereby we utilize small sections of a larger simulated quantum computer to build a library of circuits based purely on (compiled) Clifford gates.
We then show it is possible, post-training, at least for the noise model studied here, to use an ML algorithm for inference on larger circuits with higher degrees of entanglement and for non-Clifford gates without requiring further access to the quantum computer.
We do not claim the ML model deployed here is optimal for this task, only that it is able to perform the task to a useful degree of accuracy.
Further, we do not settle whether the fidelity may be predicted to arbitrary accuracy --- this is work in progress.
Additionally, testing this algorithm for other noise models, in particular for real quantum hardware, is also work in progress.

\par


In particular, we use the following two strategies for producing a theoretically inexpensive training dataset:

\vskip .5em
\noindent\textbf{Clifford to non-Clifford circuits.} The fidelity of Clifford circuits can be estimated efficiently using DFE.
The technique however does not work for general (non-Clifford) circuits.
We would like to train an algorithm using only Clifford gates to make true fidelity estimation affordable.
However, for the algorithm to generalize to non-Clifford circuits, the circuits in the training dataset must contain non-Clifford gates. 
Here this is accomplished by introducing non-Clifford gates along with their inverses into Clifford circuits in sequence --- we refer to this as the ``Clifford-reducible'' construction.
In this case, the final state wavefunction is still efficient to simulate.
We hypothesize this technique allows the ML algorithm to learn the noise characteristics of non-Clifford gates while still producing a circuit where the noiseless target state is classically accessible.

\vskip .5em
\noindent\textbf{Tiled to non-tiled circuits.} The fidelity of circuits performed on a small number of qubits can be estimated efficiently using quantum state tomography.
We leverage this to train our ML models by constructing ``tiled'' circuits by producing quantum circuits with non-overlapping segments on the qubit lattice (called ``tiles'') with no entangling gates acting across these segments.
The fidelity of the overall circuit can be computed as the product of the fidelities of the individual tiled sub-circuits.
By using circuits with different tiling structures in the training dataset, we can expose the ML algorithm to a full sampling of the local noise behaviors and allow for generalization to non-tiled circuits.


There is a feature of the tiling simulation studies in this work that requires mention: it is possible to completely remove inactive qubits.
Furthermore, resource constraints and simulation costs form a strong incentive to do so.
The noise model presented in this work contained nearest-neighbor interactions, including a cross-talk component modeled as a $ZZ$ unitary interaction, discussed in more detail in Section \ref{sec:datagen}.
As a cost-saving measure, we did not include a partial cross-talk effect when simulating tiled circuits as subsamples of the full qubit lattice.
Real quantum hardware does not allow for this removal, but in exchange it will naturally model the interactions with qubits outside the tiled sub-circuit.



\vskip .5em
\par
All of our experiments are performed using simulations with custom models. 
We utilize the \cirq \cite{cirq_developers_2021} and \texttt{Qsimcirq} software packages to simulate noiseless and noisy quantum circuits.
Demonstration experiments on hardware devices are planned for future work.


\section{Previous work}

Noise is an omnipresent feature in quantum circuits on modern hardware devices, and machine learning is a popular strategy for handling various properties of the noise. 
Multiple studies have been performed relating the domains of machine learning, quantum information, noise in quantum circuits. 

Strikis, Qin, Chen, Benjamin and Li \cite{hinsche2021learnability} studied the process of learning a statistical model to predict output distributions and used machine learning techniques to provide a strategy for quantum error compensation. 
They provided an efficient protocol for quantum noise mitigation. 
Flurin, Martin,  Hacohen-Gourgy, and Siddiqi \cite{PhysRevX.10.011006} train a neural network in real time to give a trajectory of evolution of the super-conducting qubit under certain noise models such as decoherence, etc.
Harper, Flammia and Wallman \cite{Harper_2020} give a protocol to estimate quantum noise and determine correlations in subsets of qubits on a 14-qubit superconducting quantum architecture.

Zhang et al \cite{PhysRevLett.127.130503} utilize machine learning to make predictions about quantum fidelity from a smaller set of measurements than are required for the formal calculation.
While considering a subset of the full set of possible Pauli operators they show it is possible to use a neural network to predict which of a finite set of fidelity intervals the state belongs to.
Similar to this work, Zhang et al use machine learning as a heuristic algorithm to compress information about quantum state fidelity, but they do not make predictions about circuit performance.
Rather, their work relates to improving the shot efficiency of fidelity estimation measurements given a final state wavefunction, regardless of how it is produced. 

In \cite{huangchenpreskill} Huang, Chen, and Preskill propose an efficient ML algorithm for predicting the expectation value of an operator over a process, like a quantum circuit of arbitrary depth, within a small error, given query access to the process.
Importantly, the algorithm may be trained using only random product input states and randomized Pauli measurements on the output states, but inference is possible with highly-entangled inputs, and for arbitrary expectation values.

Liu and Zhou \cite{9251243} employ a ``black-box'' model utilizing physically motivated parameters (e.g., circuit depth, the number of CNOT gates utilized, etc.) and infer a circuit reliability benchmark --- the probability of successful trial (PST, \cite{10.1145/3297858.3304007}) --- using a polynomial fit and a shallow neural-network.
The PST uses the probability (estimated with Monte Carlo) of the all-zero's bitstring running a given circuit and its inverse in sequence.
Our work differs in that our machine learning model learns an implicit representation of the noise, and we attempt to generalize training based on compartmentalized entanglement to inference tasks on circuits with broader entanglement and non-Clifford gates.
We also use a different reliability benchmark, although theirs has the advantage of being directly portable to hardware.
We will deploy an approach similar to their black-box approach as a benchmark (see the ``gate-counting regressor'' in Section \ref{sec:multimomentcircuits}).

Wang et al \cite{https://doi.org/10.48550/arxiv.2210.16724} is the closest work to this one, and debuted just before this work.
They use a graph transformer neural network (NN) model to learn the PST using both simulated and real hardware circuits.
Their model includes some explicit qubit parameters in the graph model as well as some global parameters, making it a hybrid implicit-explicit algorithm.
They additionally consider a set of structured circuits as well as random circuits.

In this paper, we use a NN to estimate the fidelity of a quantum circuit which the network has not been seen previously. 
The NN must be trained on circuits run on a specific quantum computer, but once the algorithm has been fit, our approach no longer requires access to quantum resources to make fidelity estimates.
The largest difference between this work and prior works is our focus on utilizing quantum circuits that may be simulated on classical resources with low cost for training by using ``hidden'' non-Clifford gates and compartmentalized entanglement, and extending this result to inference on circuits that are substantially harder to simulate classically.
We consider a different family of neural networks (3D convolutional networks), but consider the main innovation in this paper to be a focus on training with classically simulable circuits with a goal of extending application to beyond-classical circuits.
Here we use the exact circuit fidelity for a metric although this will not scale to large hardware circuits.
See Section \ref{sec:conclusionsandoutlook} for future plans.

\section{Methodology}

\subsection{Data generation}
\label{sec:datagen}

In this work we choose a simple but realistic noise model, adapted from \cite{z2sim}, and generate random circuits of fixed depth using the connectivity map of a square lattice quantum computer, with nearest neighbor connectivity for the 2-qubit gates.
For tiling purposes we include a variety of rectangular and square grids of varying sizes.
For benchmark scoring of our techniques we use 
$3\times 3$ and $5\times 5$ qubit square-lattice grids.
We did not engineer our dataset to create a particular distribution of fidelities.

\vskip .5em
\noindent\textbf{Circuit depth}
We consider two basic approaches.
First, we study circuits that are only one moment deep.
A moment is a set of gates that are all running in the same slice of time (functionally simultaneously) \cite{cirq_moment}.
Models trained on this type of circuit may be used to approximate multi-moment circuit fidelities by multiplying the fidelity of each moment together.
This approach has an advantage in that it may be easily extended to an arbitrary number of moments, but the disadvantage of assuming the circuit fidelity of each moment is independent of the input wavefunction.
Second, we study multi-moment circuits directly.
This model is less flexible but requires fewer assumptions about quantum circuit behavior.
Note our simulations do not include time dependent noise sources, so the multiplicative factorization is a reasonable assumption.
This is not true in general for real quantum hardware, where non-stationary noises pose serious challenges to practitioners.

\begin{figure}[h]
    \centering
    $\includegraphics[scale = 1]{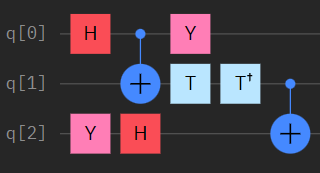}$
    \caption{Sample Clifford-reducible circuit. 
    Each vertical slice of the circuit represents a moment, or ``layer'' of gates.
    In this example we have five moments.
    }
    \label{fig:sneaky_circuit}
\end{figure}

\vskip .5em
\noindent\textbf{Clifford-reducible circuits} We construct some training circuits with both Clifford and non-Clifford gates in such a way that they are equivalent, in the noiseless case, to Clifford circuits.
This is accomplished by introducing (\gatet, $\gatet^\dagger$) pairs in sequence over pairs of moments.
These gates combine to create an effective identity operation on the qubit they are applied to.
See Figure \ref{fig:sneaky_circuit} for an illustration.
This allows for the possibility of using direct fidelity estimation to estimate the fidelity of the circuits (by removing the pairs to restore efficiently simulable behavior), which is considerably faster than tomography.

\vskip .5em
\noindent\textbf{Distribution of gates}
We consider a set of six single-qubit gates: \gatex, \gatey, \gatez, \gateh, \gatet, $\gatet ^ \dagger$ (with \gatet~and $\gatet ^ \dagger$ appearing only in multi-moment Clifford-reducible circuits); and a set of two two-qubit gates: \gatecx~(CNOT) and \gatecz. We choose gates randomly from this set to generate a circuit.
Gate placement is discussed below.

\vskip .5em
\noindent\textbf{Tiled circuits} Some training circuits for the $5\times 5$ grid are generated by generating smaller square and rectangular lattices and combining them together edge-to-edge in a process we refer to as ``tiling.'' 
Tiled circuits have separable target states.
This allows us to the estimate the fidelity of the complete circuit as a product of fidelities of the component circuits. 
This reduces computation costs for estimating fidelities for the training circuits, in comparison to non-tiled circuits, which may feature higher degrees of entanglement. 

\vskip .5em
\noindent\textbf{Noise model}
We consider two approaches to simulating noise in our quantum circuits.
The first, referred to here as the stochastic product approach, utilizes a set of models where we labeled circuits as having succeeded or failed stochastically, with failure probabilities assigned to one and two-qubit gate operations in an uncorrelated fashion.
In this approach we looped over all the elements in a circuit and computed an overall fidelity proxy by multiplying the gate success probabilities together.
We considered a set of varying gate failure probabilities, detailed in Section \ref{sec:singlemomentcircuits}.
We tested for differing failure rates as a function of location on the qubit lattice and as a function of the gate applied to test whether the ML model could discover these different patterns.
\par
In our second approach, we ran full circuit simulations using \cirq, including the custom quantum noise model mentioned above. 
Note that we study both single and multi-moment circuits with full simulation.
The custom model implements two kinds of noise: symmetric local depolarization on $1$ and $2$ qubit gates and crosstalk applied to CNOT gates.
The depolarizing operator acts upon an $n$-qubit system as follows:
$$D_n(\rho, \epsilon) = (1 - \epsilon) \rho + \frac{\epsilon}{2^n}I$$
Here, $\rho$ is the density matrix of the system and $\epsilon$ is the probability of depolarization occurring.
In \cite{z2sim}, a formulation of $D_n(\rho, \epsilon)$ as a sum over Pauli operators is given.
Our noise model applies $D_n(\rho, \epsilon _1)$ after each single qubit gate and $D_n(\rho, 10 \epsilon_1)$ after each 2-qubit gate. For our purposes, we fixed $\epsilon_1 = 0.1$ so that there is a $0.1 \%$ and $1 \%$ chance of applying local depolarization after each single-qubit and double-qubit gate respectively. 
We mode cross-talk with a unitary operator $U_{ZZ}[\zeta]$ given by: 
$$U_{ZZ}[\zeta] = exp(-i 2 \pi \zeta T \vert11\rangle \langle11\vert).$$
The parameter $T$ dictates the amount of time the crosstalk gate is applied for, and $\zeta$ determines the strength of the noise.
To apply this noise model, we insert $U_{ZZ}[\zeta]$ after every two qubit gate.
Our crosstalk strength parameters are $T = 10^{-8}$ seconds and $\zeta = 150000.$
When using full circuit simulation, because we have the full state information we use the true fidelity of the circuit as our target metric.
Note that our custom noise model does not use location-dependent or gate dependent noises (beyond the distinction between one and two-qubit gates) for reasons we will explain below.

\subsection{Circuit representation with one-hot encoding}
Our quantum circuits are represented as 4-dimensional tensors.
The first dimension encodes the moment within the circuit: $C[0]$ is the first layer, $C[1]$ is the second layer, etc.
See Figure \ref{fig:sneaky_circuit} for a pictorial representation of a circuit acting on a set of qubits through moments of gates.
The second and third dimensions are the Cartesian $x$ and $y$ coordinates of a given qubit on the square lattice respectively. 
Circuits are implemented assuming nearest-neighbor connectivity on a square lattice, so $C[n][x][y]$ gives the qubit in moment $n$ at $(x, y)$.
The fourth dimension encodes a one-hot vector where a $1$ at a particular index indicates the specific gate applied to the corresponding qubit.
The possible gates are listed in Section \ref{sec:datagen}.
For instance, $C[n][x][y][2]$ equals $1$ if there is a $Y$ gate applied to the qubit at $(x, y)$ in layer $n$.
For two-qubit gates, the one-hot index also specifies the secondary qubit that the gate acts upon. 
We convert tensor-formatted circuits to a \cirq~representation for simulation by iterating through the tensor and applying gates according to the one-hot encoding.

\subsection{Neural network architecture and training}
\label{sec:nnarch0}

For single-moment circuit regressor neural networks, we chose to use a 2D Locally Connected Network \cite{lcnkeras} without any form of regularization or input normalization.
For multi-moment circuit regressor neural networks, we used a 3D Convolutional Neural Network (CNN) with no regularization and no input or batch normalization. 
Additional details on training are provided in Appendix \ref{sec:nntraindetails}.
The architectures are provided in Appendix \ref{sec:nnarch}.

\section{Experiments and results}

\subsection{Single moment circuits}
\label{sec:singlemomentcircuits}

\subsubsection{Stochastic product model noise}
\label{sec:adhocprob}

We employed a variety of uncorrelated stochastic gate fidelity models:
\begin{enumerate}
    \item In the first approach, single qubit gates are assigned success probabilities of 0.99, and 2-qubit gates are assigned success probabilities of 0.95.
    This means we assign a 99\% (95\%) chance of each one (two) qubit gate operating as intended.
    Otherwise we label the result as an error state.
    \item In the next approach, each gate has a unique success probability between 1 and 0.94: $I = 1$, $, Y = 0.98$, $Z = 0.97$, $H = 0.96$, $CX = 0.95$, $CZ = 0.94$.
    Interpretation of these values is as in the previous case.
    \item In the next approach, we set the success probability of the outer qubits in the lattice to be higher than that of inner qubits (to simulate a bulk crosstalk-like effect).
    Specifically, all the single qubit gates have a success probability of 0.99 and all the two qubit gates have a success probability of 0.95 if they are associated with edge qubits.
    All the single qubit gates have a success probability of 0.97 and all the two qubit gates have a success probability of 0.93 if they are associated with qubits situated off the edge of the lattice.
    \item In the final ad-hoc approach, each gate is assigned a unique success probability between 1 and 0.94: $I = 1$, $, Y = 0.98$, $Z = 0.97$, $H = 0.96$, $CX = 0.95$, $CZ = 0.94$, and to simulate crosstalk, we assign an additional failure ``penalty'' (0.94) to all qubits adjacent to a gate. This means for a single qubit gate, we check all 4 of its neighboring qubits and multiply the success probability by 0.94 everytime there is a gate associated with the neighboring qubit. More specifically, given a gate on the lattice such that three of its neighboring qubits have a gate, then the success probability is multiplied by $(0.94)^3$.
    Similarly, for two qubit gates we check all six of its nearest neighbors and multiply by 0.94 each time a gate is encountered.
\end{enumerate}

\begin{figure}[htbp]
    \centering
    \begin{subfigure}[t]{0.2\textwidth}
        $\includegraphics[scale = 0.2]{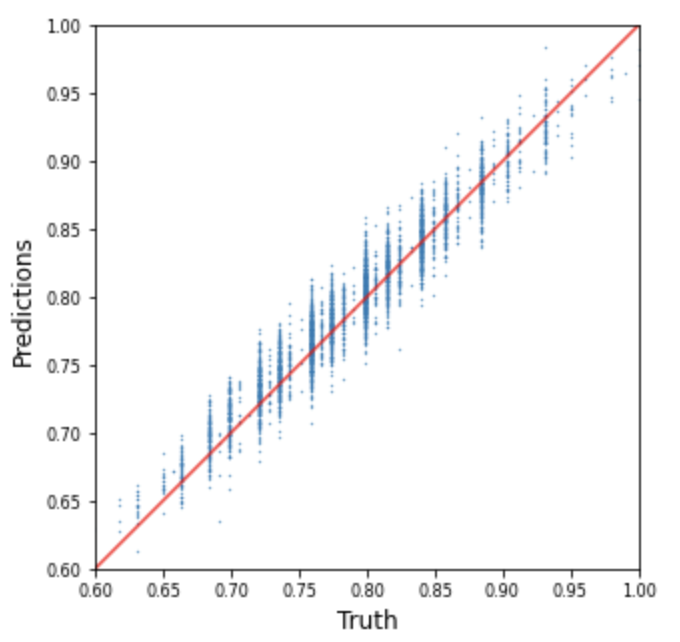}$
        \caption{Fidelity model 1.}
        \label{fig:sm_custom_fidelity_scat_1}  
\end{subfigure}
    \begin{subfigure}[t]{0.2\textwidth}
        $\includegraphics[scale = 0.2]{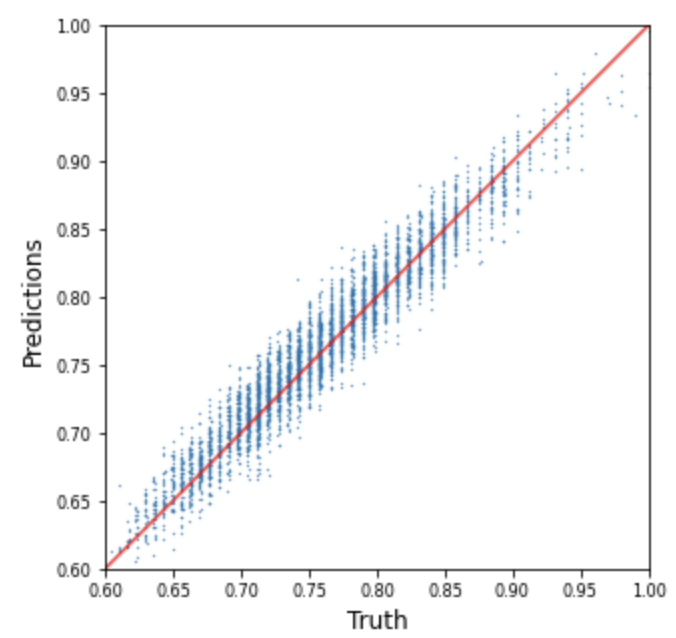}$
        \caption{Fidelity model 2.}
        \label{fig:sm_custom_fidelity_scat_2}  
    \end{subfigure}
    \begin{subfigure}[t]{0.2\textwidth}
        $\includegraphics[scale = 0.2]{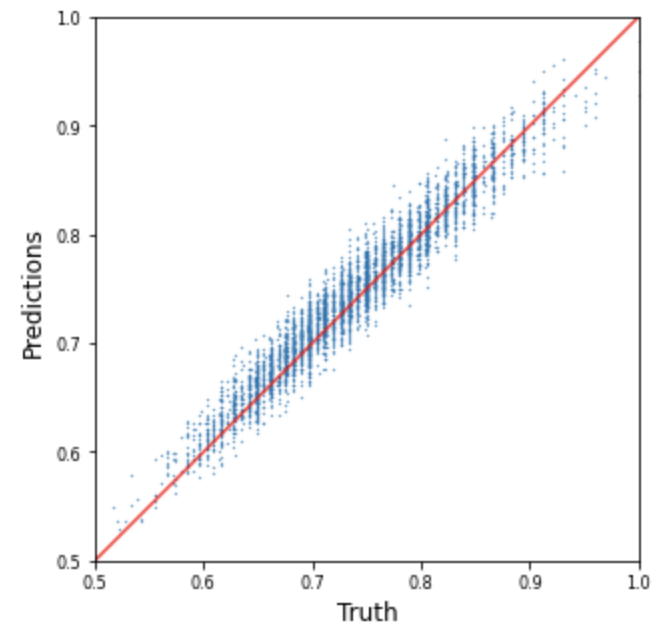}$
        \caption{Fidelity model 3.}
        \label{fig:sm_custom_fidelity_scat_3}  
    \end{subfigure}
    \begin{subfigure}[t]{0.2\textwidth}
        $\includegraphics[scale = 0.2]{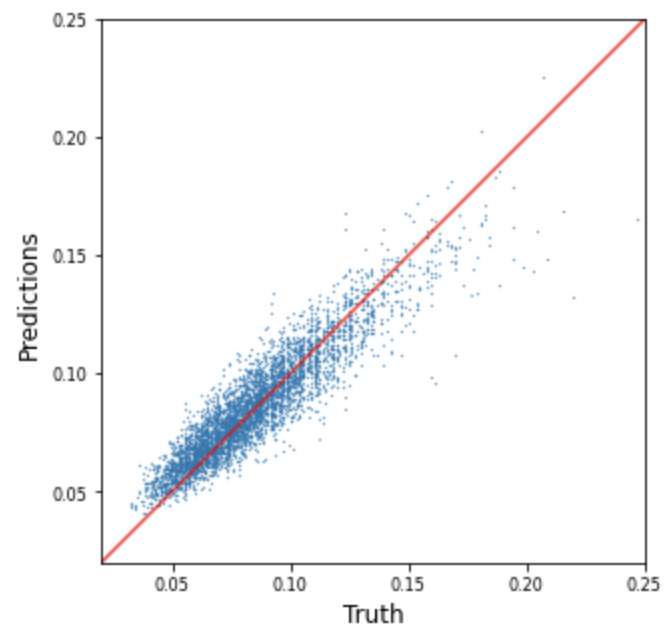}$
        \caption{Fidelity model 4.}
        \label{fig:sm_custom_fidelity_scat_4}  
    \end{subfigure}
    \caption{A 2D locally-connected model's performance on single-moment circuits implemented on a $3 \times 3$ qubit lattice with stochastic product model noises.}
    \label{fig:custom_nm}
\end{figure}

A locally-connected NN model was able to learn all four of these ad-hoc failure models and accurately predict success probability for $3 \times 3$ single moment circuits, as shown in Figure \ref{fig:custom_nm}.
Given this success, for simplicity we chose to use a uniform noise model (no variations for different qubits in the chip, or different gates) when performing full circuit simulations for multi-moment circuits.

\subsubsection{Circuit simulations with custom noise}
\label{sec:singlemomentsim}

\par
\begin{figure}[h]
    \centering
    $\includegraphics[scale = 0.5]{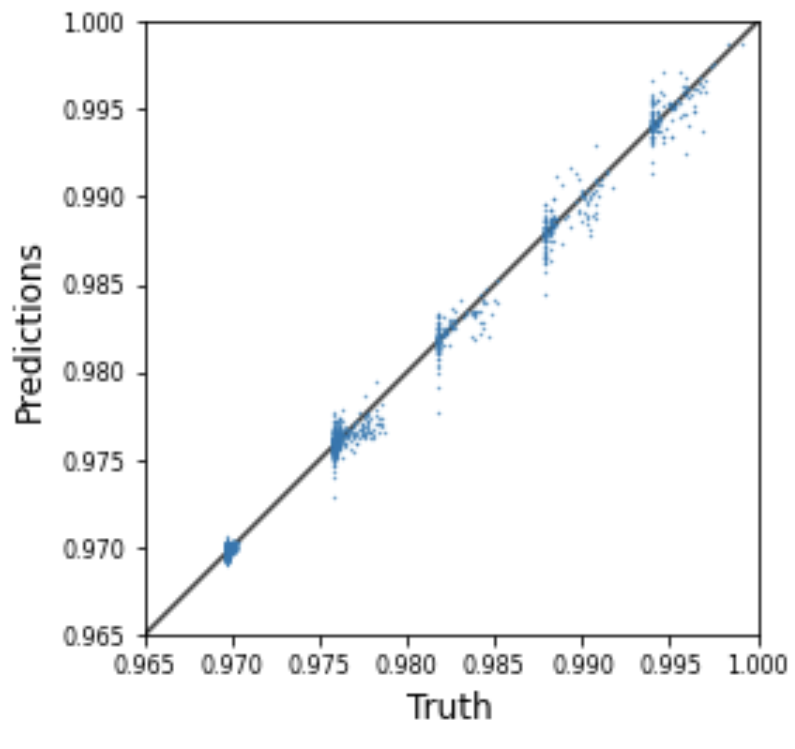}$
    \caption{Fidelity estimates for a 2D locally-connected NN model trained on 1500 $3 \times 3$ single-moment Clifford circuits and tested on 1000 $3 \times 3$ single-moment Clifford circuits. The structure in the distribution of values is related to the integer number of possible gates, and is dominated by the number of two-qubit gates, which are noisier in our model (and on most real quantum computers).
    }
    \label{fig:sm_cirq_3x3}
\end{figure}

We next swapped stochastic product models for a noise model incorporating depolarization and crosstalk implemented in a full circuit simulation using \cirq.
We also shifted from the stochastic product model of fidelity metrics to using $F(\rho, \sigma) = (tr \sqrt{\sqrt{\rho}\sigma \sqrt{\rho}})^2$ where $\rho, \sigma$ are density matrices.
We trained a locally-connected network on $3000$ simulated single-moment Clifford circuits on a $3 \times 3$ qubit lattice, and found a linear correlation between the true and predicted fidelities of a similar test dataset, as shown in Figure \ref{fig:sm_cirq_3x3}. 

\subsection{Multi-moment circuits}
\label{sec:multimomentcircuits}

Our primary benchmark for multi-moment circuits is a linear regression model that carries location-specific parameters for fidelity penalties for one and two-qubit gates.
We chose this model as a good proxy for the ``practitioner's benchmark'', which is to count one and two-qubit gates and assign an estimated fidelity penalty for each. 
We extended the model slightly by making it-location aware --- in other words, each different qubit and coupler will have an independent penalty in the fit.

We compare this benchmark, referred to as the ``gate-counting regressor'' model to the NN described in Appendix \ref{sec:nnarch}, with architecture parameters specified in Table \ref{tab:3dcnn}.
Moving forward, we shorten our description of the gate-counting regressor model to simply the ``regressor'' model even though both the NN and the linear model are regressors estimating the fidelity.

Our comparison spans three levels of increasing complexity.
In all cases we \emph{train} the estimators using Clifford-reducible, tiled circuits.
Recall these circuits are constructed such that non-Clifford gates appear only in pairs that sequence to the identity operation, and could therefore be removed at compile time for efficient classical simulation.
Furthermore, tiled circuits restrict the area of entanglement across the total lattice to again allow for inexpensive classical simulation of the final state (factorized) wavefunction.
Tiling was also a practical requirement in this work for noisy circuit simulation.
Even though our noise model is relatively simple, it is expensive to generate large simulated datasets on $5 \times 5$ lattices on the commercial cloud computing nodes available to us.

\subsubsection{Performance evaluation}
\label{sec:perfeval}

\begin{figure}[h]
 \centering
 \includegraphics[width=.85\textwidth]{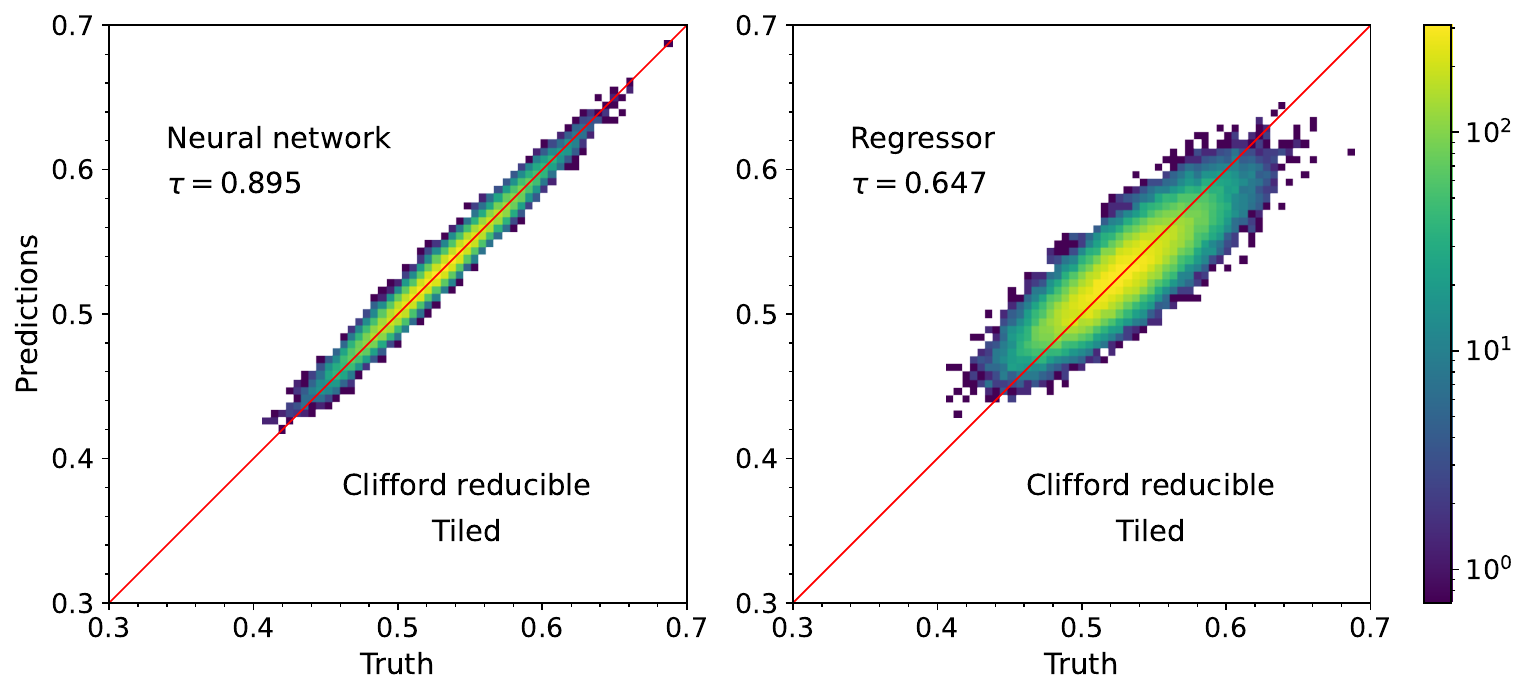}\\
 \includegraphics[width=.85\textwidth]{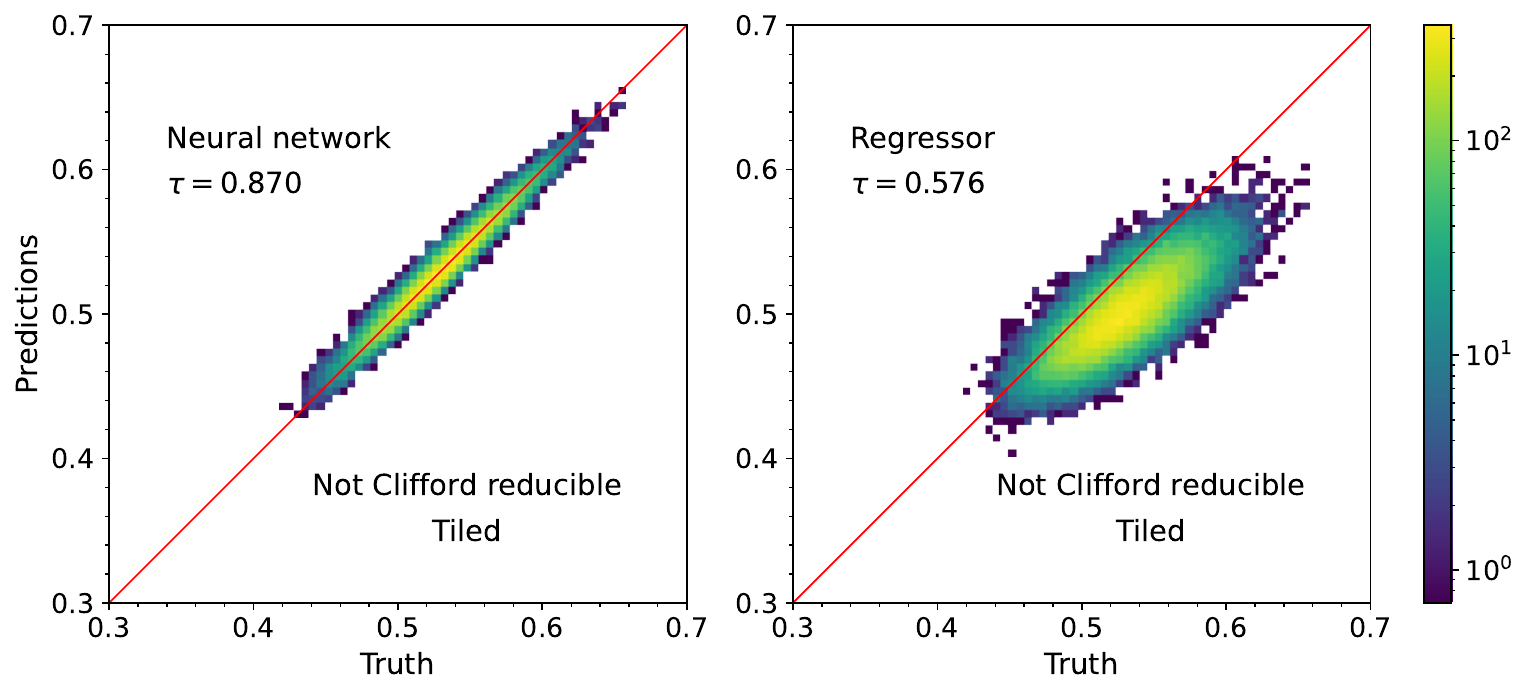}\\
 \includegraphics[width=.85\textwidth]{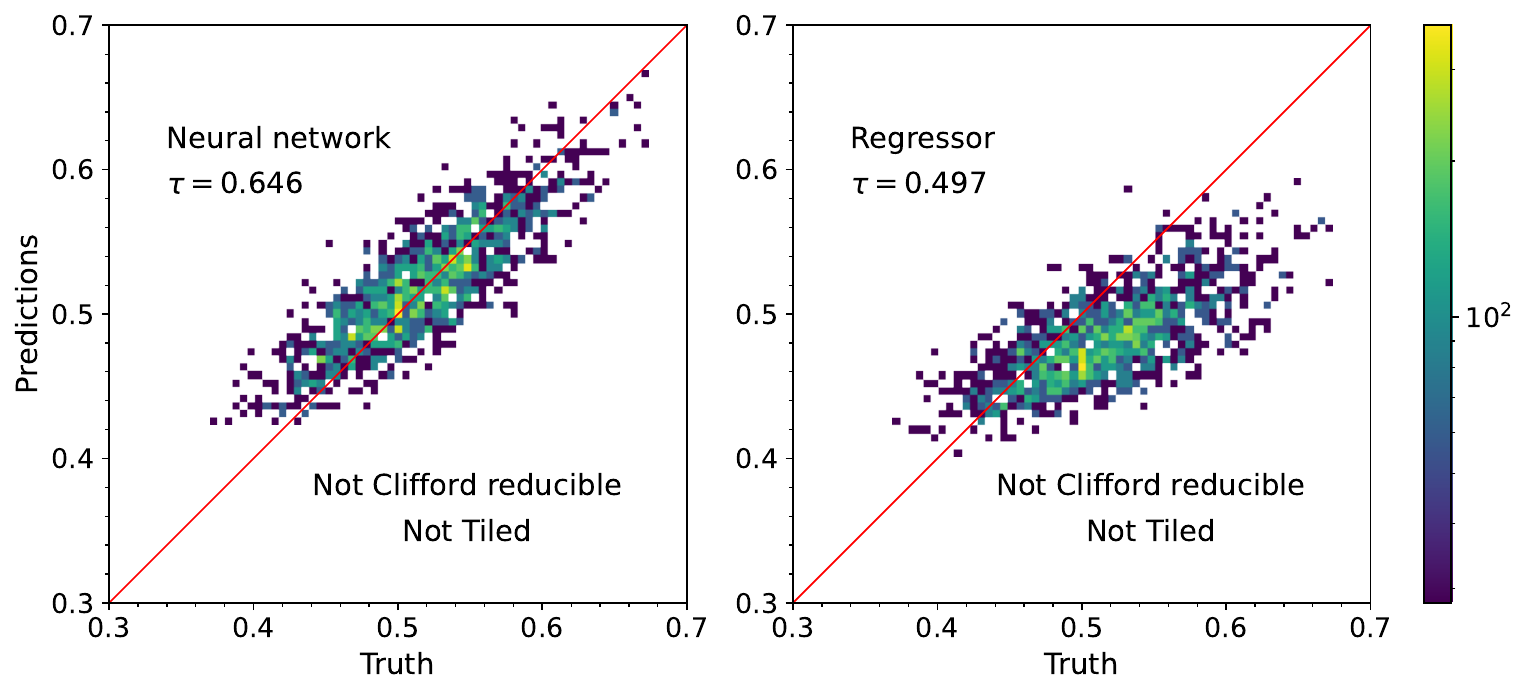}
 \captionof{figure}{Normalized heatmaps showing the predicted vs true fidelities for a neural network estimator and a gate counting regressor, both trained on tiled, Clifford-reducible circuits. The top, middle, and bottom panels correspond to testing the estimators on a) tiled, Clifford-reducible circuits, b) tiled, non-Clifford-reducible circuits, and c) non-tiled, non-Clifford-reducible circuits, respectively. The plots also show the Kendall-tau coefficient between the predicted and true fidelities in each case.}
 \label{fig:estimator_benchmarks}
\end{figure}

Qualitative estimator benchmarks at the \emph{inference stage} are summarized in Figure \ref{fig:estimator_benchmarks}.
We first consider a case where we benchmark the estimators on Clifford-reducible, tiled circuits.
Both estimators show predictive power for the fidelity, although the more sophisticated neural network model shows a tighter spread in the relationship between true and predicted fidelities.
We next consider tiled, non-Clifford-reducible circuits.
Again both estimators show predictive power, but the simpler regressor model shows a systematic bias in its estimations, along with a broader spread in the true vs predicted fidelities.
We finally consider non-tiled, non-Clifford-reducible circuits.
Both estimators continue to show predictive power, and similar spreads in the true vs predicted fidelity distributions, but the simpler regressor continues to show a bias.

Loosely inspired by the receiver operating characteristic (ROC) curve, we studied an additional metric that attempted to capture the ``usefulness'' of the estimators for deciding whether or not the estimator prediction for fidelity may be relied on.
We hypothesize one major use case of models like these would be to decide if an experiment on hardware is likely enough to succeed to be worth investing quantum resources.
For the question of whether a given circuit has a fidelity at least $t$, the true positive rate and true negative rate of a given predictor are given by
\begin{align}
 \mathrm{TPR}(t) = \mathrm{Pr}(F_p \geq t~\vert~F_t \geq t)\,,\\
 \mathrm{TNR}(t) = \mathrm{Pr}(F_p < t~\vert~F_t < t)\,,
\end{align}
where $F_p$ is the predicted fidelity, $F_t$ is the true fidelity, and $t$ is a given threshold value for the fidelity. Here we will use half the sum of $\mathrm{TPR}$ and $\mathrm{TNR}$ as a metric to evaluate the performance of the different fidelity estimators.
It produces a quality curve over the range of fidelities of interest somewhat similar to a receiver operating characteristic (ROC) curve in that the area under the curve may be used to establish higher quality performance over a range of operation.
A random guessing strategy will give a flat score across fidelity thresholds at 0.5, while a perfect classifier will give a flat score across fidelity thresholds at 1.0.
We prefer this metric to mean-squared error (MSE) to summarize performance because MSE can be misleading when integrated over the full test dataset, and for quantum circuits our score directly addresses the question --- ``is this circuit worth running on this hardware?'' 

\begin{figure}
 \centering
 \includegraphics[width=0.66\columnwidth]{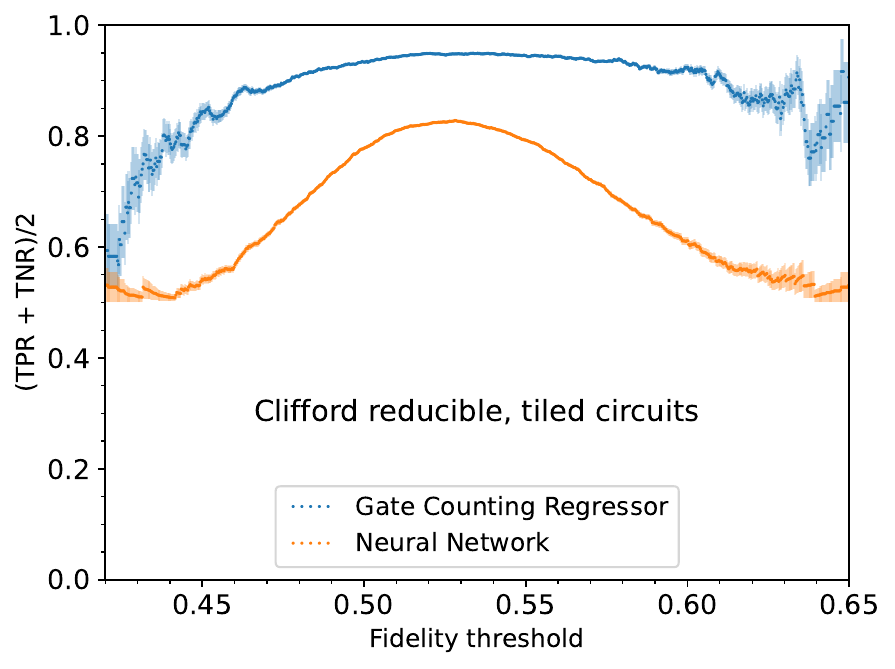}\\
 \includegraphics[width=0.66\columnwidth]{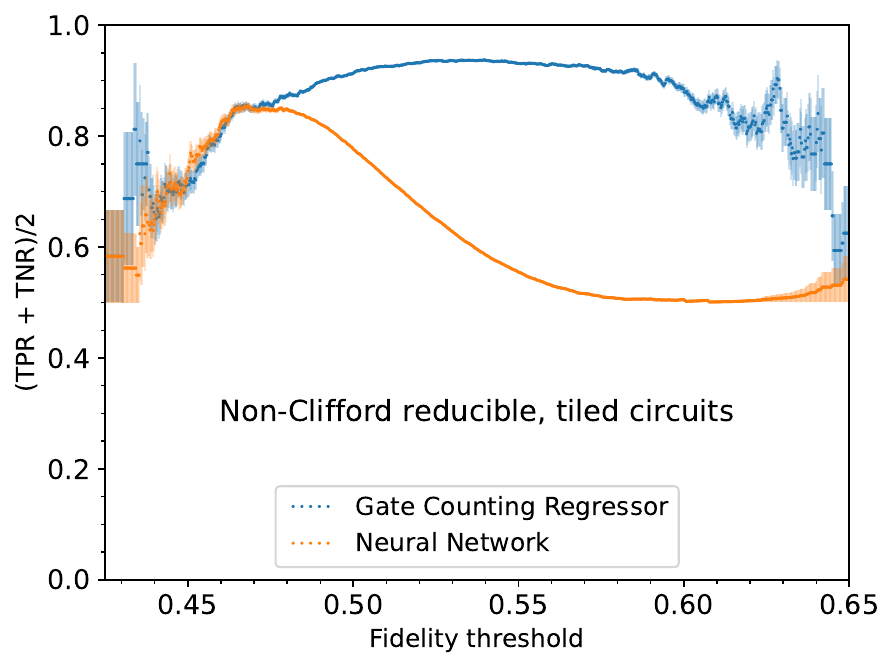}\\
 \includegraphics[width=0.66\columnwidth]{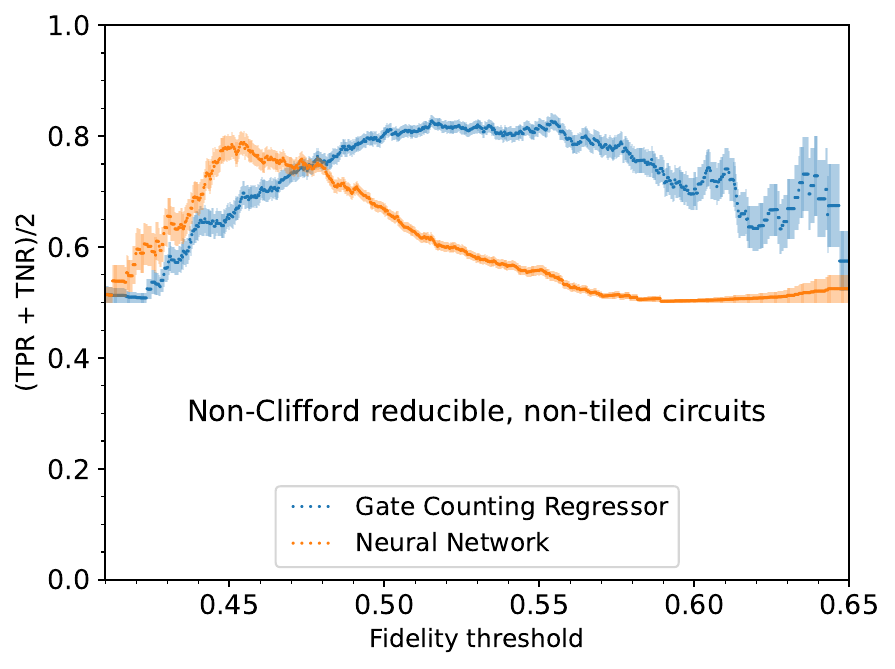}
 \caption{Figures showing the metric $(\mathrm{TPR}(t) + \mathrm{TNR}(t))/2$ for various fidelity thresholds $t$, for the neural network estimator and the gate counting estimator, both trained on tilded, Clifford-reducible circuits. The top, middle, and bottom panels correspond to testing the estimators on a) tiled, Clifford-reducible circuits, b) tiled, non-Clifford-reducible circuits, and c) non-tiled, non-Clifford-reducible circuits, respectively. $\mathrm{TPR}$, $\mathrm{TNR}$ and their uncertainties were estimated using Wilson's formulas for binomial-proportion confidence intervals.}
 \label{fig:shyamsundar_scores}
\end{figure}

We show the score as a function of fidelity threshold for both estimators in Figure \ref{fig:shyamsundar_scores}.
Note that while a better performing model will have higher scores, it is also the case that the shape of the performance curves is impacted by the distribution of fidelities in the underlying dataset.
Furthermore, while our score is inspired by the ROC curve, it is not really correct to compare the areas under these curves as an absolute standard given the influence of the underlying distributions.
Nonetheless, we feel these curves provide an additional useful comparison between estimators.

\section{Conclusions and outlook}
\label{sec:conclusionsandoutlook}

In this work we proposed a method to perform heuristic estimation of circuit fidelity using machine learning and provided an empirical demonstration of a classical ML algorithm trained on efficient-to-simulate circuits, but able to run inference on circuits that feature exponential simulation scaling costs.
While this demonstration is empirical, and we must restrict our claims to the noise model studied here, we believe result motivates further theoretical investigation into the expressive power of machine learning for capturing the behavior of complex systems.

We do not claim the method presented here will be successful on real quantum hardware at this time because the features of real quantum noise are different from our simulated noise model in ways that may be significant.
For example, noise on real quantum hardware is often non-stationary and non-local.
Further, we do not claim the specific ML model evaluated here is optimal for this task, and we note the model utilized in this work is specific to a square-lattice qubit connectivity.

We emphasize this method is one tool among many for studying and predicting the performance of a noisy quantum computer.
While this method requires access to a quantum computer during training, and there are good reasons to believe it will require re-training when faced with non-stationary noise models, once a model is obtained it allows for the estimation of performance metrics with no further use of the quantum computer.
This is potentially valuable in compilers or when deciding whether certain experiments are worth testing on a real QPU, where access time may be especially precious.
 
In future work we will test this approach on real quantum hardware for performance metric prediction.
On hardware, full process tomography is prohibitively expensive for even a moderate number of qubits, making other metrics like MCFE and PST attractive proxies.
Additionally, here we consider circuits of fixed depth, but we are studying extensions to allow for circuits of varying depth.

\section{Acknowledgements}

We thank E. Peters for useful code, comments, and instructions in adapting the quantum noise model employed in this work.
We also thank members of the QMLQCF collaboration, in particular A. del Gado, J. Duarte, and J. R. Vlimant for enlightening discussion about this work.
\par
This document was prepared using the resources of the Fermi National Accelerator Laboratory (Fermilab), a U.S. Department of Energy (DOE), Office of Science, HEP User Facility. 
Fermilab is managed by Fermi Research Alliance, LLC (FRA), acting under Contract No. DE-AC02-07CH11359.
R.K., G. P., and P. S. were supported for this work by the DOE/HEP QuantISED program grant ``HEP Machine Learning and Optimization Go Quantum,'' identification number 0000240323, and P. S. was additionally supported by the DOE/HEP QuantISED program grant ``QCCFP-QMLQCF Consortium,'' identification number DE-SC0019219.

\newpage
\appendix

\section{Neural network training details}
\label{sec:nntraindetails}

For our single-moment circuit network regressors, the loss was computed via the mean-squared error (MSE), and the model was optimized using stochastic gradient descent (SGD) with learning-rate $= 0.01$, momentum $= 0.9$, and a Nesterov Accelerated Gradient. 
During training, both the training loss and validation loss converged to $O(10 ^ {-8})$.
The architecture is provided in Appendix \ref{sec:singlemomentnetstruc}.

For multi-moment circuit network regressors, we used a 3D Convolutional Neural Network (CNN) to predict fidelities.
During training, the loss was computed by mean squared error (MSE), and the model was optimized using stochastic gradient descent (SGD) with learning-rate $= 0.01$, momentum $= 0.9$, and a Nesterov Accelerated Gradient.
For example, the loss and validation loss of the 3D CNN for 12-moment $3 \times 3$ Clifford circuits converges to $O(10 ^ {-4})$, as shown in Figure \ref{fig:CNN_3x3_loss}.
The architecture is provided in Appendix \ref{sec:multimomentnetstruc}.

\begin{figure}[h]
    \centering
    $\includegraphics[scale = 0.6]{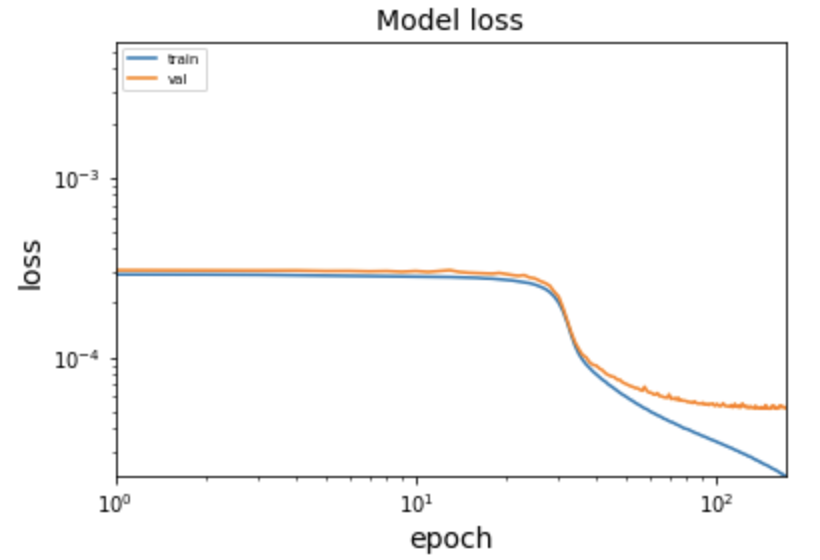}$
    \caption{Loss of 3D CNN trained on 15,000 12-moment $3 \times 3$ Clifford circuits.
    }
    \label{fig:CNN_3x3_loss}
\end{figure}

\section{Neural network architectures}
\label{sec:nnarch}
We designed, by-hand, a pair of neural network (NN) architectures for the single and multi-moment problems.

\subsection{Single-moment network structure}
\label{sec:singlemomentnetstruc}

The layer-by-layer details for the single-moment NN model --- utilizing 2D locally-connected layers --- is
given in Table \ref{tab:LC}.

\begin{table}[] 
\centering
\caption{The structure summary for our model for single-moment 13-channel $3\times 3$ circuits. We use a locally connected network. The zero-padding 2D layer (ZP2D) pads the input with zeros for use in subsequent layers.
The first layer re-shapes the input to account for the one-hot gate encoding. The first locally connected 2D layer (LC2D) uses 16 channels with a kernel size of (2,2) and stride length of (1,1). The subsequent LC2D layers use 16, 16, and 1 channel respectively (kernel size of (1,1) each time).}
\begin{tabular}{p{2.25cm} p{2.5cm} p{2cm}} 
Layer & Output Shape & Params    \\ \hline \hline 
ZP2D & (5, 5, 13) & 0 \\ \hline
LC2D & (4, 4, 16) & 13568 \\ \hline 
LC2D & (4, 4, 16) & 4352 \\ \hline 
LC2D & (4, 4, 16) & 4352 \\ \hline 
LC2D & (4, 4, 1) & 272 \\ \hline 
Flatten & (16) & 0 \\ \hline 
Dense & (64) & 1088 \\ \hline 
Dense & (64) & 4160 \\ \hline 
Dense & (1) & 65 \\ \hline \hline
\end{tabular}
\newline
\\
Total params: 27,857
\label{tab:LC} 
\end{table}

\subsection{Mutli-moment network structure}
\label{sec:multimomentnetstruc}

The layer-by-layer details for the multi-moment NN model --- utilizing 3D convolutional layers --- is
given in Table \ref{tab:3dcnn}.

\begin{table}[] 
\centering
\caption{The structure summary for our multi-moment model. 
We use three successive 3D convolutional layers with 50 channels, kernel size of (4,4,4), and stride length of (1,1,1). The input to the convolutional layers are padded with 0s so that the outputs have the same 3D dimensions as the input.}
\begin{tabular}{p{1.8cm} p{3cm} p{1.8cm}} 
Layer & Output Shape & Params    \\ \hline \hline 
Conv3D & (12, 5, 5, 50) & 48050 \\ \hline
Conv3D & (12, 5, 5, 50) & 160050 \\ \hline 
Conv3D & (12, 5, 5, 50) & 160050 \\ \hline 
Flatten & (15000) & 0 \\ \hline 
Dense & (500) & 7500500 \\ \hline 
Dense & (500) & 250500 \\ \hline 
Dense & (500) & 250500 \\ \hline 
Dense & (500) & 250500 \\ \hline 
Dense & (50) & 25050 \\ \hline 
Dense & (1) & 51 \\ \hline 
Dense & (1) & 2 \\ \hline \hline
\end{tabular}
\newline
\\
Total params: 8,645,253
\label{tab:3dcnn} 
\end{table}



\begin{thebibliography}{18}
\providecommand{\natexlab}[1]{#1}
\providecommand{\url}[1]{{#1}}
\providecommand{\urlprefix}{URL }
\providecommand{\doi}[1]{\url{https://doi.org/#1}}
\providecommand{\eprint}[2][]{\url{#2}}
 \bibcommenthead

\bibitem[{Boixo et~al(2018)Boixo, Isakov, Smelyanskiy, Babbush, Ding, Jiang,
  Bremner, Martinis, and Neven}]{Boixo2018}
Boixo S, Isakov SV, Smelyanskiy VN, et~al (2018) Characterizing quantum
  supremacy in near-term devices. Nature Physics 14(6):595--600.
  \doi{10.1038/s41567-018-0124-x},
  \urlprefix\url{https://doi.org/10.1038/s41567-018-0124-x}

\bibitem[{{Cirq Developers}(2021)}]{cirq_developers_2021}
{Cirq Developers} (2021) Cirq. \doi{10.5281/zenodo.5182845},
  \urlprefix\url{https://doi.org/10.5281/zenodo.5182845}

\bibitem[{{Cirq Developers}(2022)}]{cirq_moment}
{Cirq Developers} (2022) Cirq documentation.
  \urlprefix\url{https://quantumai.google/cirq/build/circuits}

\bibitem[{Flammia and Liu(2011)}]{PhysRevLett.106.230501}
Flammia ST, Liu YK (2011) Direct fidelity estimation from few pauli
  measurements. Phys Rev Lett 106:230,501.
  \doi{10.1103/PhysRevLett.106.230501},
  \urlprefix\url{https://link.aps.org/doi/10.1103/PhysRevLett.106.230501}

\bibitem[{Flurin et~al(2020)Flurin, Martin, Hacohen-Gourgy, and
  Siddiqi}]{PhysRevX.10.011006}
Flurin E, Martin LS, Hacohen-Gourgy S, et~al (2020) Using a recurrent neural
  network to reconstruct quantum dynamics of a superconducting qubit from
  physical observations. Phys Rev X 10:011,006.
  \doi{10.1103/PhysRevX.10.011006},
  \urlprefix\url{https://link.aps.org/doi/10.1103/PhysRevX.10.011006}

\bibitem[{Gustafson et~al(2021)Gustafson, Holzman, Kowalkowski, Lamm, Li,
  Perdue, Boixo, Isakov, Martin, Thomson, Heidweiller, Beall, Ganahl, Vidal,
  and Peters}]{z2sim}
Gustafson E, Holzman B, Kowalkowski J, et~al (2021) Large scale multi-node
  simulations of $\mathbb{Z}_2$ gauge theory quantum circuits using google
  cloud platform. \doi{10.48550/ARXIV.2110.07482},
  \urlprefix\url{https://arxiv.org/abs/2110.07482}

\bibitem[{Harper et~al(2020)Harper, Flammia, and Wallman}]{Harper_2020}
Harper R, Flammia ST, Wallman JJ (2020) Efficient learning of quantum noise.
  Nature Physics 16(12):1184--1188. \doi{10.1038/s41567-020-0992-8},
  \urlprefix\url{https://doi.org/10.1038%2Fs41567-020-0992-8}

\bibitem[{Hinsche et~al(2021)Hinsche, Ioannou, Nietner, Haferkamp, Quek,
  Hangleiter, Seifert, Eisert, and Sweke}]{hinsche2021learnability}
Hinsche M, Ioannou M, Nietner A, et~al (2021) Learnability of the output
  distributions of local quantum circuits. \eprint{2110.05517}

\bibitem[{Huang et~al(2022)Huang, Chen, and Preskill}]{huangchenpreskill}
Huang HY, Chen S, Preskill J (2022) Learning to predict arbitrary quantum
  processes. \doi{10.48550/ARXIV.2210.14894},
  \urlprefix\url{https://arxiv.org/abs/2210.14894}

\bibitem[{Keras(????)}]{lcnkeras}
Keras (????) \url{https://keras.io/api/layers/locally_connected_layers/},
  accessed: 2022-09-26

\bibitem[{Liu and Zhou(2020)}]{9251243}
Liu J, Zhou H (2020) Reliability modeling of nisq- era quantum computers. In:
  2020 IEEE International Symposium on Workload Characterization (IISWC). IEEE
  Computer Society, Los Alamitos, CA, USA, pp 94--105,
  \doi{10.1109/IISWC50251.2020.00018},
  \urlprefix\url{https://doi.ieeecomputersociety.org/10.1109/IISWC50251.2020.00018}

\bibitem[{Poyatos et~al(1997)Poyatos, Cirac, and Zoller}]{PhysRevLett.78.390}
Poyatos JF, Cirac JI, Zoller P (1997) Complete characterization of a quantum
  process: The two-bit quantum gate. Phys Rev Lett 78:390--393.
  \doi{10.1103/PhysRevLett.78.390},
  \urlprefix\url{https://link.aps.org/doi/10.1103/PhysRevLett.78.390}

\bibitem[{Proctor et~al(2022{\natexlab{a}})Proctor, Rudinger, Young, Nielsen,
  and Blume-Kohout}]{Proctor2022}
Proctor T, Rudinger K, Young K, et~al (2022{\natexlab{a}}) Measuring the
  capabilities of quantum computers. Nature Physics 18(1):75--79.
  \doi{10.1038/s41567-021-01409-7},
  \urlprefix\url{https://doi.org/10.1038/s41567-021-01409-7}

\bibitem[{Proctor et~al(2022{\natexlab{b}})Proctor, Seritan, Nielsen, Rudinger,
  Young, Blume-Kohout, and Sarovar}]{Proctor2022EstablishingTI}
Proctor T, Seritan S, Nielsen E, et~al (2022{\natexlab{b}}) Establishing trust
  in quantum computations. \doi{10.48550/ARXIV.2204.07568},
  \urlprefix\url{https://arxiv.org/abs/2204.07568}

\bibitem[{da~Silva et~al(2011)da~Silva, Landon-Cardinal, and
  Poulin}]{PhysRevLett.107.210404}
da~Silva MP, Landon-Cardinal O, Poulin D (2011) Practical characterization of
  quantum devices without tomography. Phys Rev Lett 107:210,404.
  \doi{10.1103/PhysRevLett.107.210404},
  \urlprefix\url{https://link.aps.org/doi/10.1103/PhysRevLett.107.210404}

\bibitem[{Tannu and Qureshi(2019)}]{10.1145/3297858.3304007}
Tannu SS, Qureshi MK (2019) Not all qubits are created equal: A case for
  variability-aware policies for nisq-era quantum computers. In: Proceedings of
  the Twenty-Fourth International Conference on Architectural Support for
  Programming Languages and Operating Systems. Association for Computing
  Machinery, New York, NY, USA, ASPLOS '19, p 987–999,
  \doi{10.1145/3297858.3304007},
  \urlprefix\url{https://doi.org/10.1145/3297858.3304007}

\bibitem[{Wang et~al(2022)Wang, Liu, Cheng, Liang, Gu, Li, Ding, Jiang, Shi,
  Qian, Pan, Chong, and Han}]{https://doi.org/10.48550/arxiv.2210.16724}
Wang H, Liu P, Cheng J, et~al (2022) Quest: Graph transformer for quantum
  circuit reliability estimation. \doi{10.48550/ARXIV.2210.16724},
  \urlprefix\url{https://arxiv.org/abs/2210.16724}

\bibitem[{Zhang et~al(2021)Zhang, Luo, Wen, Feng, Pang, Luo, and
  Zhou}]{PhysRevLett.127.130503}
Zhang X, Luo M, Wen Z, et~al (2021) Direct fidelity estimation of quantum
  states using machine learning. Phys Rev Lett 127:130,503.
  \doi{10.1103/PhysRevLett.127.130503},
  \urlprefix\url{https://link.aps.org/doi/10.1103/PhysRevLett.127.130503}

\end{thebibliography}


\end{document}